\documentclass{aa}  

\usepackage{graphicx}
\usepackage{hyperref}
\usepackage{txfonts,textcomp}

\usepackage[normalem]{ulem}

\begin{document} 

   \title{When LAMOST meets {\textit{Gaia}} DR3}

   \subtitle{Exploring the metallicity of open clusters}

 \author{R. Zhang
          \inst{1}
          \and
 Guo-Jian Wang\inst{5,6}
 \and
 Yuxi(Lucy) Lu\inst{7}
 \and
 Sufen Guo\inst{2,1}
 \and 
  S. Lucatello\inst{3,10}
 \and
 Xiaoting Fu\inst{11}
  \and
  Haifeng Wang\inst{4}
  \and
 Luqian Wang\inst{1}
 \and
 Schiappacasse-Ulloa, J\inst{3,4}
  \and
 Jianxing Chen\inst{8,9}
 \and
 Zhanwen Han\inst{1}
 }
   \institute{ Yunnan Observatories, Chinese Academy of Sciences, Kunming 650216, P.R. China\\
   \email{ruyuan.zhang@ynao.ac.cn;yl4331@columbia.edu;guosufen@ynao.ac.cn}
   \and
     School of Physical Science and Technology, Xinjiang University, Urumqi 830046, People's Republic of China 
   \and
   INAF-Osservatorio Astronomico di Padova, vicolo Osservatorio 5, 35122, Padova, Italy
   \and 
  Dipartimento di Fisica e Astronomia ``Galileo Galilei", Universita di Padova, Vicolo Osservatorio 3, I-35122, Padova, Italy
   \and 
   Department of Physics, Stellenbosch University, Matieland 7602, South Africa
  \and           
   National Institute for Theoretical and Computational Sciences (NITheCS), South Africa
\and
  Astrophysics Department, American Museum of Natural History, Central Park West, New York, 10024, New York, USA.
\and
Institute for Frontiers in Astronomy and Astrophysics, Beijing Normal University, Beijing 102206, P. R. China
\and
School of Physics and Astronomy, Beijing Normal University, Beijing 100875, P. R. China
  \and
  Institute for Advanced Studies, Technische Universität München, Lichtenbergstraße 2 a, 85748 Garching bei München
  \and
  Purple Mountain Observatory, Chinese Academy of Sciences, Nanjing, 210023, China
}
   \date{Received }

 
  \abstract
   {
Open clusters (OCs) {are excellent probes} as their age and abundance can be tightly constrained, allowing us to explore the distribution of metallicity and composition across the disk of the Milky Way. By conducting a comprehensive analysis of the metallicity of OCs, we can obtain valuable information about the history of their chemical enrichment. Moreover, by observing stars in different regions of the Milky Way, we can identify significant spatial structures in their chemical composition and abundance. This enables us to understand stellar birth radii through chemical tagging. Nevertheless, it remains challenging to infer the original positions of OCs using current data alone.
}
   {The aim of this  study is to investigate the distribution of metallicity in the solar neighborhood using a large dataset from {\textit{Gaia}} DR3 combined with LAMOST spectra. 
   With accurate ages and metallicity measurements, we can determine birth radii for the stars and attempt to understand their migration pattern.}
   {
We chose a total of 1131 OCs within 3 Kpc of the Sun from the {\textit{Gaia}} DR3 and LAMOST DR8 low-resolution spectral database (R=1800). We used an artificial neural network to correct the LAMOST data by incorporating high-resolution spectral data from GALAH DR3 (R=28000). The average metallicity of the OCs was determined based on the reliable [Fe/H] values for their members. We then examined the distribution of metallicity across different regions within the Galaxy and inferred birth radii of the OCs from their age and metallicity.
}
   {
The correction method presented here can partially eliminate the systematic offset for {LAMOST} data. We discuss the metallicity trend as a function of Galactocentric distance and the guiding radii. We also compare these observational results with those from chemo-dynamic simulations. Values derived from observational metallicity data are slightly lower than predicted values when the uncertainties are not considered. However, the metallicity gradients are consistent with previous calculations. Finally, we investigated the birthplace of OCs and find hints that the majority of OCs near the Sun have migrated from the outer Galactic disk. 

}
   {}


   \maketitle
%

\section{Introduction}

Galactic archaeology seeks to unravel the past of the  Milky Way and to construct a timeline of its evolution. By analyzing observational data for stars, we aim to understand the Galaxy's environment during its formation and subsequent evolution. The kinematics, age, and chemical composition of the stars allow us to derive tight constraints on their birthplace. 
Stars inherit their chemical composition from the interstellar medium (ISM) where they form, including elements produced through various processes such as supernovae, stellar winds,  and neutron star mergers. This enables the identification of stars with similar chemical patterns as originating from the same molecular cloud at specific moments in time, a technique known as chemical tagging \citep[see e.g.,][]{Freeman02}. While molecular clouds within the same galaxy may exhibit unique compositions, accurately tracing stars to their original ISM requires precise chemical analysis \citep{Desilva06}. The method of weak chemical labeling proposed by Vanessa Hill \citep[see e.g.,][]{gilmore22}, combined with age and dynamic information (e.g., \citealt{wang2020a, wang2020c} and references therein), offers a promising approach to exploring the original locations of stars, overcoming challenges in analyzing large samples.


Determining the precise ages of stars is vital for constructing a timeline of the formation of the Milky Way. However, age measurement, especially for older stars, poses challenges and entails greater uncertainties. In addition to traditional isochrone fitting, age determination relies on combining photometric data across multiple bands with spectrum analysis using sky survey data \citep{sanders18,queiroz18,kordopatis23,2023arXiv231017196W}. Furthermore, a new method based on asteroseismology provides a more reliable approach to determining stellar age \citep{chaplin13,chaplin14,chaplin20,anders23}. Another promising method is the use of "chemical clocks", which rely on tracking the changes in element abundances within stars as a result of nuclear reactions occurring in their interiors. These reactions can either increase or decrease the abundances of certain elements as the star ages \citep{hayden22,viscasillas22,moya22}. The integration of stellar age with chemical abundance could be used as a tool for establishing a comprehensive event calendar.


Open clusters (OCs) are stellar groups held together by gravity and formed from a single molecular cloud. They are predominantly young ---typically less than 1 Gyr   \citep{fujii16}--- 
with the oldest existing for around 8 Gyr before dissipation \citep{salaris04,kharchenko13}. 
The age, distance, metallicity, and composition of OCs are relatively easy to determine
compared to those of field stars because OCs consist of stars with similar ages and chemical compositions. The spatial concentration of cluster members also facilitates accurate distance measurements using methods such as main sequence fitting and isochrone fitting. 
Furthermore, the stars in an OC formed within the same ISM. This simplifies the process of determining their metallicity and chemical composition through spectroscopic analysis.
Mostly found in the Milky Way disk, OCs share metallicities of approximately the solar value, and are distributed across the galactic thin disk. Over time, OCs move away from their birth location via radial migration \citep{sellwood02}.
Studies reveal a correlation between the ages of OCs and their spatial distribution, with older clusters situated farther from the Galactic plane, influenced by vertical heating mechanisms driven by Galactic structures such as warps and minor mergers \citep[see e.g.,][]{vanden80,friel95,cantat20,marting14,mackereth19,ting19,sharma21}. These dynamic processes, involving radial migration and heating, can elevate OCs to higher vertical positions, showcasing the intricate interplay between objects and the Galactic structure \citep{sellwood02,bird12,minchev10,minchev11,minchev13}.


Open clusters offer a means to study the chemical, structural, and dynamic evolution of the Galactic disk across various positions in the Galaxy, and the [Fe/H] trend is crucial for constraining chemical evolution models and for the construction of a timeline of Galactic events. Various studies \citep{reddy16,genovali14,casamiquela19,donor20,cantat20,netopil22,myers22} have analyzed high-quality data for OC samples and stars in order to use these objects  as tracers to detect the  behavior and gradient of  chemical composition across galactic space. 
The metallicity gradient derived from OCs typically has a value of around -0.060 dex kpc$^{-1}$, but flattens toward the outer disk \citep{friel10,donor20,myers22}. \cite{myers22} compared the metallicity trend observed in APOGEE DR17 with results predicted from chemo-dynamical models, discovering that observational results exhibit a systematic bias for older OCs when compared to predictions. These authors suggested that this bias may be attributed to the migration of OCs.

The work of \cite{minchev10a} and \cite{minchev18} focuses on mechanisms of disk migration. Many studies use large survey samples to further explore galactic disk migration. \cite{chen20} and \cite{zhang21} selected samples of OCs based on the High-resolution spectra (HRS) data provided by \cite{netopil16}, and found evidence of migration. These samples consist of three types of migration modes, and these authors conducted their analysis of migration based on the distribution of metallicity. \cite{viscasillas23} investigated the interaction between field stars and OCs and galactic disk migration using {\textit{Gaia}}-DR3 data.


Currently, a few thousand OCs have been identified using photometric data from large Sky surveys, and less than 10\% of these have been the target of high-resolution spectroscopic observations prior to recent advancements. 
There are several major surveys ---such as {\textit{Gaia}}-ESO, APOGEE, and GALAH--- \footnote{https://www.galah-survey.org/;GALAH has released its third set of data, which provides high-resolution (R=28000) spectra and covers four wavelength ranges: 4713-4903, 5648-5873, 6478-6737, and 7585-7887 \AA.}  that have collected information about OCs using high-resolution spectral data, although the sample sizes are limited. {\textit{Gaia}} Data Release 3 (DR3) has provided information on approximately 5.6 million stars through its Radial Velocity Spectrometer (RVS) \citep{cropper18,katz23}. Using the General Stellar Parametriser spectroscopy module (GSP-Spec), stellar atmospheric parameters and composition abundance were determined \citep{recio23}. Compared to {\textit{Gaia}} DR2, parallax and proper motion accuracies have improved by 30\% and 100\%, respectively \citep{Marton23}, and the 
{\textit{Gaia}} DR3 target magnitude range spans from stars brighter than 13.6 in G band to faint stars of around 25 in G band {\citep{riello21,Gaia23}}

The Large Sky Area Multi-Object Fiber Spectroscopic Telescope (LAMOST)\footnote{https://www.lamost.org/} is located in Xinglong, China. It has a diameter of 4 meters and consists of two spectral modules. LAMOST is the only instrument capable of obtaining data for the outer-disk OCs. Thanks to its powerful capability, LAMOST has observed millions of stars, covering a wide range of disk positions \citep{deng12}. This dataset includes the latest {\textit{Gaia}} DR3 photometric data and features an extensive number of OC samples with metallicity measurements, significantly expanding the sample size for outer-disk OCs.

This paper is organized as follows: Section \ref{sec2} presents the sample selection and the reference data, providing a detailed comparison of the radial velocity (RV) of the samples and the correction of metallicity and $\alpha$-elements using an ANN. In Sect. \ref{sec3}, we discuss the differences between our results and those gathered from the literature. Section \ref{sec4} examines the distribution of metallicity across various Galactic spatial regions and the migration patterns of OCs. Finally, we summarize our findings in Sect. \ref{sec5}.


\section{Data and samples}\label{sec2}

 {\textit{Gaia}} DR3 data can be used to improve the identification of OC membership and can provide more accurate distances for distant clusters. In their catalog, \cite{hunt23} reported over 7000 clusters and stellar associations, including 2378 newly identified clusters and 134 globular clusters spanning the entire sky. Additionally, \cite{cavallo23} developed an ANN model using a new feature-extraction technique called the QuadTree algorithm \citep{schiap23}. This model was used to re-estimate the parameters of OCs based on Hunt's catalog. However, the determination of metallicity in this study relied on photometric data, and the metallicities obtained are hence of modest accuracy, as shown in Fig. 7 of \cite{cavallo23}. Although {\textit{Gaia}} has expanded the number of OCs with available chemical abundance data, such information is limited to objects within a relatively bright magnitude limit within which {\textit{Gaia}} RVS can effectively collect information. 
In their work, \cite{viscasillas23} used the {\textit{Gaia}} DR3 samples with chemical abundance data from GSP-Spec. These authors selected clusters and field stars that covered the Galactocentric range of 7.5-9 kpc and investigated the radial migration of objects. However, the availability of samples in the outer range of the disk is limited.

LAMOST is a highly effective telescope, and is able to observe approximately 4000 targets per exposure. The low-resolution module has a resolution of 1800 and covers the wavelength range of 3700  to 9000 \AA. The medium-resolution module has a resolution of 7500. In those modules, LAMOST uses two cameras: the blue camera captures spectra from 4950 to 5350 \AA, while the red camera captures spectra from 6300 to 6800 \AA. LAMOST has already released its DR8 dataset for the international community, which includes low- and medium-resolution spectra for over 5 million objects \citep{liu20}. The LAMOST stellar parameter pipeline (LASP) analyzes LAMOST's low-resolution spectra and determines key atmospheric parameters. Spectral analysis involves extracting 1D spectra post-data reduction, including corrections and combining sub-exposures.
It also involves detecting initial parameters using correlation function interpolation \cite[CFI][]{Du2012} and deriving final parameters with ULySS; \citep{koleva2009, wy11, wu2011} by minimizing $\chi^2$\footnote{https://www.lamost.org/dr8/v2.0/}. The details of the spectral template used to measure atmospheric parameters are summarized in the work of \cite{bai21}.

Additionally, LAMOST has entered the phase of medium-resolution observations, leading to an increase of approximately 2 million stellar samples. However, at present, there are no publicly accessible pipelines specifically developed for the analysis of LAMOST medium-resolution spectra \citep{luo15,ren15}, and therefore we do not analyze these here. In the present study, a total of 1131 OCs were selected, which includes a total of 8640 stars. Of these, more than 1020 OCs have less than three stars, while the remaining OCs typically contain five stars.

Recently, \cite{cavallo23} undertook a new analysis of stellar associations made up of more than ten stars using the catalog compiled by \cite{hunt23}. The authors focused on approximately 6300 OCs, and used an ANN to determine their fundamental parameters. These parameters were derived from {\textit{Gaia}} and 2MASS data, including age, extinction, distance, and metallicity. 
We selected the samples for the present study by matching the positions of objects within 3 arcsecs \citep{cui12} between the LAMOST catalog and the catalog of \cite{cavallo23}. Furthermore, we cross-checked the citation ID number from LAMOST with the {\textit{Gaia}} DR3 source ID to ensure consistency. Of the identified stars, over 8000 were found to belong to approximately 1000 OCs listed in Cavallo's catalog.




\subsection{Radial velocity}

The RVs employed in this section are derived from the low-resolution spectra (LRS) obtained from the LAMOST Data Release 8 catalog. LASP uses cross-correlation to identify spectral classes and determine redshifts or RVs. To tackle the challenges of detection, a low-order polynomial is used to align spectra. Currently, SDSS templates are being used; however, LAMOST templates will be updated after the first data are released \citep{luo15}. 
These velocities are influenced by the S/N and are determined through the optimal fit of $\chi^2$. However, it is important to note that RV measurements derived from LRS are affected by  significant
uncertainties, which warrant attention. There are noticeable differences between RVs obtained from LRS and those derived from high-resolution spectra \citep{fu22}. As the members of the OCs are identified from the literature, the process of identifying membership is not included in the present study. The comparison results are taken directly from the LASP DR8 catalog, where the uncertainty associated with the LAMOST RVs peaks around 5 km/s.
We noted the radial-velocity scatter in a single OC and used the LASP outcome, considering different signal-to-noise ratios  (S/Ns). 
 The RV dispersion values obtained depend on S/N and spectral type, as indicated by \cite{bai21}. The corresponding standard deviation ranges for FGK stars are approximately 7.67-13.18 km/s, 8.60-6.59 km/s, 5.30-6.19 km/s, and 3.93-5.17 km/s for S/N values of 5-10, 10-20, 20-40, and  > 40, respectively.
The RV of an OC becomes more uniform as the S/N increases. The accuracy of RV measurements for individual stars relies on their S/N, with higher S/N measurements being considered more reliable. Stars with significant RV measurement discrepancies are flagged as potential binaries and will be excluded from further selection unless validated as single stars. Ultimately, the RV of an OC is calculated by averaging the RVs of its  individual members.

\begin{figure}
   \centering
   \includegraphics[width=10cm]{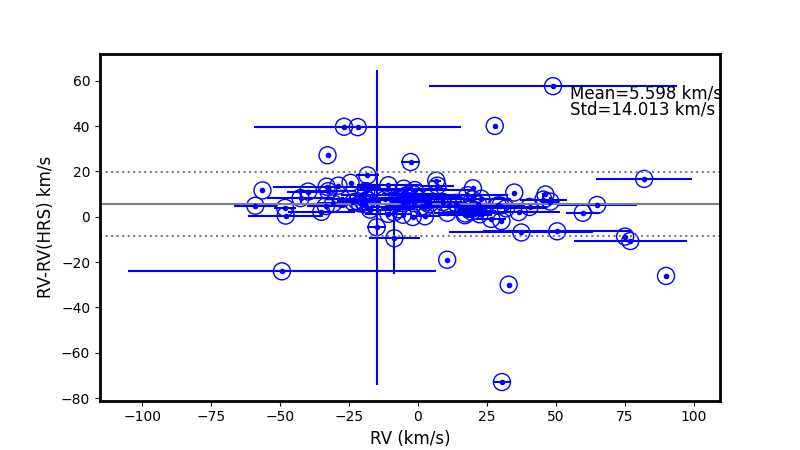}
      \caption{
Difference in RVs between 101 cross-matched targets from previous works, including those by \citet{casali19}, \citet{casamiquela16}, \citet{spina21}, \citet{frasca19}, \citet{casali20}, \citet{dorazi20}, \citet{paperI}, and \citet{carrera22}, and the current study is shown. The gray lines indicate the mean offset value of 5.598 km/s with a standard deviation of 14.013 km/s for this comparison. The x-axis represents the RV of the present study, incorporating the error for the overlapping sample.}

   \label{figRV2}
   \end{figure}

The RVs used for comparison are derived from high-resolution spectra obtained through various projects, including the Gaia-ESO Survey (GES), Open Clusters Chemical Abundances from Spanish Observatories (OCCASO), GALactic Archaeology with HERMES (GALAH), Stellar Populations and Abundances (SPA), and One Star to Tag Them All (OSTTA) datasets \citep{casali19, casamiquela16, spina21, frasca19, casali20, dorazi20, paperI, carrera22}. Additionally, we include the sample studied by \cite{donor20} using APOGEE DR16 in our RV comparison.
In this study, to ensure data quality and an adequate number of samples for comparison, we imposed a minimum S/N of 30 in the final data cut. Figure \ref{figRV2} displays the RV distribution per OC and the offset between the HRS result and our work. On average, there is a difference of approximately 5.59 km $s^{-1}$, with a scatter of 14.01 km $s^{-1}$. \cite{tsantaki22} assessed the differences in RV measurements between LAMOST and other surveys (APOGEE, GALAH, RAVE), finding that the RV bias between LAMOST and high-resolution spectra is between 4 and 5 km $s^{-1}$ for spectra with a S/N of greater than 30. This value is slightly lower than but comparable to our findings.


   \begin{figure}
   \centering
   \includegraphics[width=10cm]{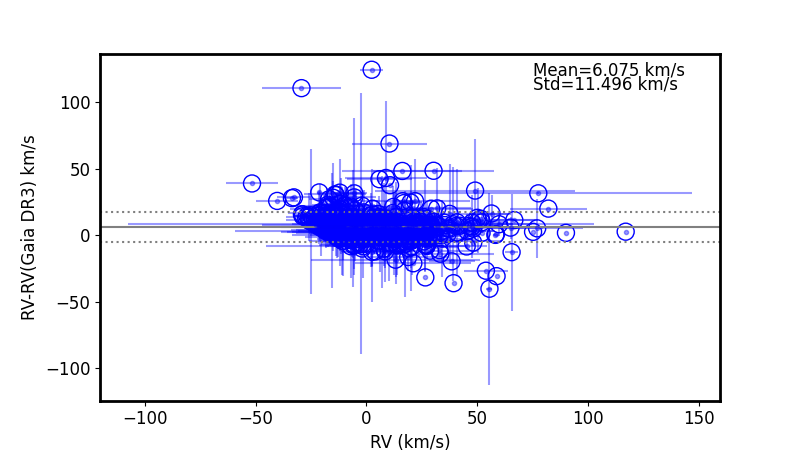}
      \caption{
Difference distribution in RVs for 595 cross-matching targets between \textit{{Gaia}} DR3 and the present study. The mean value of the offset is 6.075 with a standard deviation of 11.496 km $s^{-1}$ as indicated by the gray lines in the plot. The x-axis represents the LAMOST RV and the error of the overlapped sample. The y-axis includes the RV uncertainties from \textit{{Gaia}} DR3.}

   \label{figRV}
   \end{figure}

\begin{table*}[t]
\setlength{\tabcolsep}{1mm}
\begin{center}
\caption{Radial velocities for the OCs studied here compared to those from the literature}
\begin{tabular}{l|c|c|r|r|c|c|r|r}
\hline\hline
  Cluster        &     RA     &   Dec.      & RV(LRS)  & eRV(LRS) & RV(HRS) & eRV(HRS)&RV({\textit{Gaia}} DR3) & eRV({\textit{Gaia}} DR3)  \\
                 & (J2000)    & (J2000)   & (km/s) & (km/s) & (km/s) & (km/s) & (km/s) &(km/s) \\
\hline
\object{ADS 16795} & 23:30:22.12 & 58:33:12.29 & -20.01 & -- & -- & -- & -- &--\\
\object{ASCC 100} & 19:01:36.74 & 33:36:03.40 & -21.895 & 4.1649 & -- & -- & -14.395 & 4.1649 \\
\object{ASCC 101} & 19:13:22.73 & 36:21:35.64 & -22.94 & -- & -- & -- & -- & -- \\
\object{ASCC 105} & 19:41:53.18 & 27:23:09.94 & -18.44875 & 3.2471 & -- & -- & --& --  \\
\object{ASCC 108} & 19:53:32.93 & 39:20:33.36 & -14.962 & 3.4953 & -19.5642 & 69.4883 & 13.175& --  \\
\object{ASCC 11} & 03:32:12.10 & 44:50:26.76 & -18.3893 & 11.0881 & -10.94 & 0.0 & --& --  \\
\object{ASCC 113} & 21:11:51.02 & 38:32:43.09 & -11.6078 & 6.7277 & -- & -- & -5.395 & 6.7277 \\
\object{ASCC 12} & 04:49:40.94 & 41:42:51.48 & -14.7071 & 18.7118 & -- & -- & -14.2625 & 18.7118 \\
\object{ASCC 123} & 22:42:15.05 & 54:11:56.09 & -32.76 & -- & -5.6 & 0.0 & --& --  \\
\object{ASCC 128} & 23:20:38.75 & 54:34:24.10 & -18.6789 & 6.9543 & -- & -- & -14.76 & 6.9543 \\
\hline
\end{tabular}
\tablefoot{The uncertainties resulting from the RV measurements obtained from HRS have not been accounted for in some existing literature. Columns 7 and 8 list the OC RVs from \textit{Gaia} DR3. The full table is available at the CDS.}
\label{tab-prop}
\end{center}
\end{table*}

We find a greater scatter between the RVs of the OCs studied here compared with other studies;  this is verified again in this section. For example, the difference in RV between FSR 0951 and Berkeley 32 exceeds 50 km/s. The present study only includes two stars from FSR 0951, which increases the uncertainty in the RV measurements. Berkeley 23, an old OC, has six members, but only two stars have RV measurements that can be compared with those from the literature.
Most of the large offsets in Fig. \ref{figRV2} correspond to clusters that contain only one or two member stars, such as Collinder 107, Alessi 62, Berkeley 19, Berkeley 31, Berkeley 17, SFR 0667, ASCC 123, and UPK 422. This highlights the importance of considering the number of stars within OCs when making comparisons.

{\textit{Gaia}} DR3 provides the stellar RV for more than 300,000 stars, which include objects found in 595 of the OCs in our sample. Figure \ref{figRV} shows the RVs of {\textit{Gaia}} DR3 compared with the RVs for the sample of OCs studied in the present paper. The systematic offset is around 6.075 km $s^{-1}$, but  the significant dispersion in the RV residuals between the two databases may be attributed to RV measurement errors and unverified binaries.

\subsection{Metallicity determination}
This section focuses on controlling the quality of the data. The analysis presented here was carried out using the LAMOST official pipeline LASP, which has now published eight data releases. The LAMOST General catalog contains the atmospheric parameters (effective temperature, surface gravity, and metallicity) of F, G, and K stars. Currently, this catalog has published spectral data for over six million stars. To ensure high quality, as mentioned above, only the final results with a spectral S/N of greater than 30 were selected. However, the metallicity values show a systematic bias, especially for low-temperature dwarfs with an effective temperature of below 5000 K and surface gravity higher than 4.3 dex \citep{fu22,soubiran22}. Similar properties are derived for these stars when using the data-driven DD-Payne method \citep{xiang19}, which relies on observed spectra for training. 
The spectral analysis of cool stars presents larger uncertainties. This is because the analysis becomes more challenging as the stars get cooler, mainly due to blending
\citep{casali20}, which leads to underestimation of metallicity in the LAMOST data. This could be inherited from the determination template mentioned in Sect. \ref{sec2}.  
To tackle this issue, we employed a calibration technique that involves adjusting the metallicity abundance. This was done using an empirical correction relationship derived from wide binary stars. Furthermore, the metallicity of cooler stars was
calibrated by referencing their hotter companions \citep{niu23}. 
However, based on the current data, the average standard deviation of [Fe/H] for the chosen OCs is approximately 0.16 dex. Additionally, over 10\% of the clusters exhibit a metallicity dispersion of greater than 0.25 dex, surpassing the desired chemical abundance precision of 0.05 dex \citep{Freeman02}. In this section, we aim to correct the stellar metallicity using a similar method to that used by \cite{spina21}. We use an ANN to build a model that corrects the iron abundance for individual stars.

In the remaining samples, \cite{soubiran22} evaluated the accuracy of the measurements of [Fe/H] as provided by multiple surveys, including APOGEE,GALAH, RAVE, LAMOST, SEGUE, and {\textit{Gaia}}-ESO. These authors used samples of
 OCs to investigate internal uncertainties. They found that the residuals of [Fe/H] changed by between -0.3 and -0.15 dex  within a temperature range from 4000 K to 6500 K. In this section, we correct the data from LAMOST using the data from the GALAH survey. The GALAH survey aims to investigate the origin of the Milky Way, including its dynamic processes and chemical composition. The survey provides 30 different chemical abundance measurements for around 600,000 stars  \citep{buder21}.

For our training samples, we selected around 30,000 targets that were observed in both the LAMOST and GALAH surveys. When choosing these samples, we avoided stars that are too cool and for which the spectra are dominated by molecular lines\citep{jofre19}, and those that are too hot and lack metal lines for analysis. Our selection criteria align with those mentioned by \cite{spina21}: the effective temperature must fall within the range of 4000 K and 7000 K, the surface gravity must be > 0 dex, the metallicity must be between -1 and 0.5 dex, 
and the microturbulence should be smaller than 2.5 km/s. Additional criteria include   $\lvert$ $T_{\textrm{eff,LAMOST}}$-$T_{\textrm{eff,GALAH}}$ $\rvert$ < 150 K,$\lvert$ log g$_{\textrm{LAMOST}}$-log g$_{\textrm{GALAH}}$ $\rvert$ < 0.3 dex, and $\lvert$ [Fe/H]$_{\textrm{LAMOST}}$-[Fe/H]$_{\textrm{GALAH}}$ $\rvert$ < 0.1 dex.

\begin{figure}
   \centering
   \includegraphics[width=8.5cm]{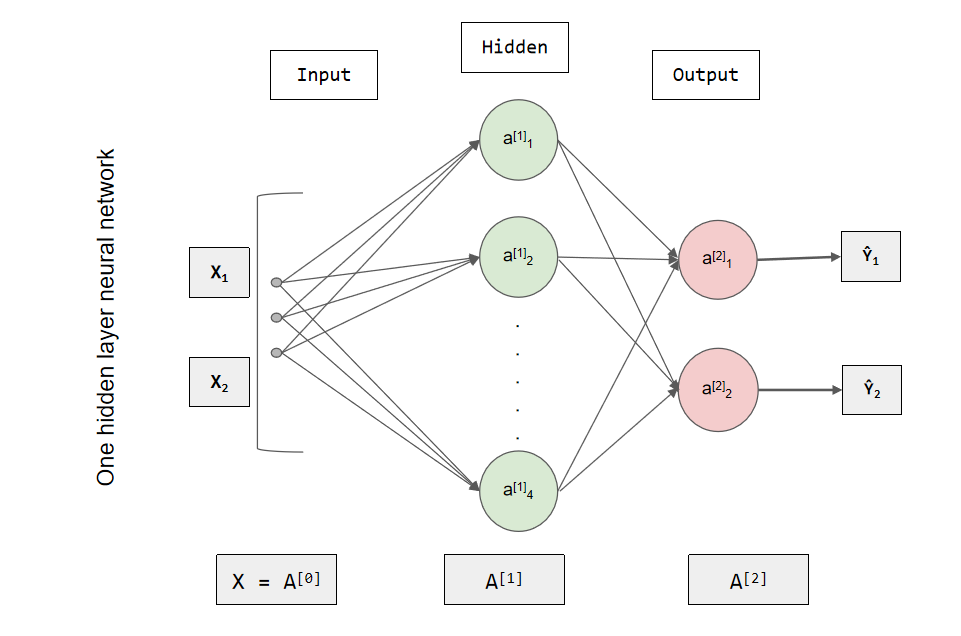}
    \caption{Structure of the ANN used in this work. The number of neurons in the hidden layer is 128.}\label{fignn}
\end{figure}

 We used our ANN to achieve the correction. An ANN is a computational model that takes inspiration from biological neural networks. The general structure of an ANN is made up of an input layer, several hidden layers, and an output layer. An ANN is trained on a data set to learn a mapping between the input and the output. In the training process, the hyperparameters (weight $w$ and bias $b$) of the ANN are optimized by minimizing a loss function. For the correction here, we trained an ANN to learn a mapping between the LAMOST data and the GALAH data. To this end, we fed the ANN model with the LAMOST metallicity and [Fe/H] uncertainties, expecting to obtain the corresponding calibration values. We built and trained  the ANN model using the code CoLFI\footnote{https://github.com/Guo-Jian-Wang/colfi}, which is a framework for likelihood-free inference with neural density estimators \citep{wanggj2020, wanggj2022, wanggj2023}. The ANN model adopted here contains a hidden layer with 128 neurons (see Fig. \ref{fignn}). We used the rectified linear unit (ReLU, \citet{relu}) as the nonlinear function, and the least absolute
deviation as the loss function.

\begin{figure}
   \centering
   \includegraphics[width=0.5\textwidth]{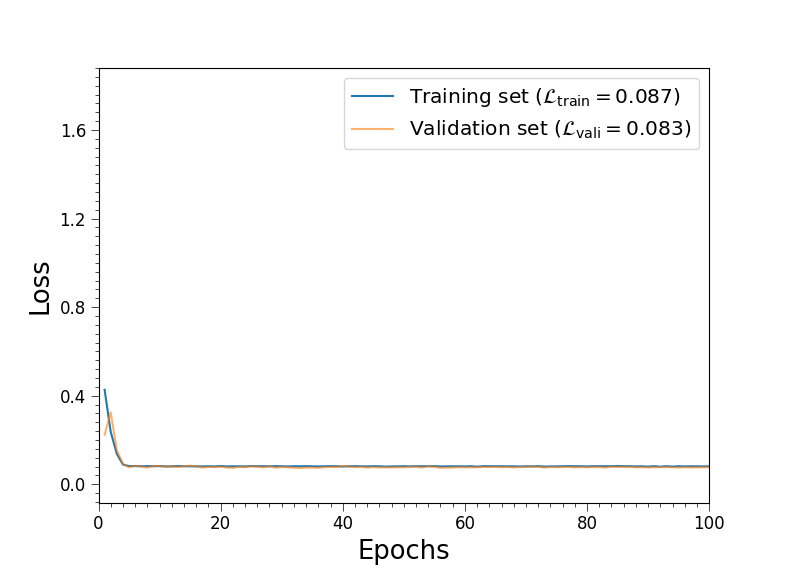}
      \caption{Loss function of the training and validation sets.}
   \label{fig_Loss}
   \end{figure}

\begin{figure*}
   \centering
   \includegraphics[height=7cm]{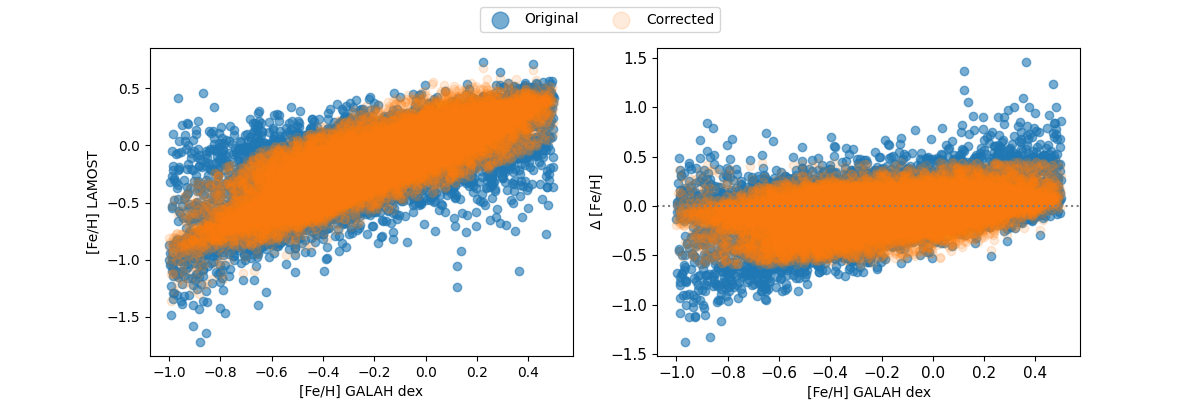}
      \caption{Comparison of Metallicity Measurements from LAMOST DR8 and GALAH.The right panel shows the iron relation for 30000 targets. The y-axis represents the metallicity from LAMOST dr8, while the x-axis represents the corresponding value from GALAH. The left plot displays the metallicity residuals of LAMOST and GALAH, both before and after correction. The blue dots represent the original data selected from LAMOST DR8, while the orange dot represents the LAMOST result after correction.}
   \label{fig_homo}
   \end{figure*}

The training set, validation set, and test set for the ANN model were divided using the traditional fraction of 6:2:2. The loss functions of the training and validation sets are depicted in Fig. \ref{fig_Loss}, indicating the absence of overfitting. Next, we used the well-trained ANN model to correct the data from LAMOST. The corrected data are presented in Fig. \ref{fig_homo}, which shows a greater concentration around the relational function. The correction is particularly effective for stars with significant differences in [Fe/H] between the two databases.


\subsection{Open cluster metallicity}

The accuracy of OC parameters can be enhanced by examining the star that shows the greatest scatter. Furthermore, the uncertainties can be reduced by averaging the results obtained from multiple stars. 
We aim to estimate the metallicity of OCs using the method described in Sect. 2.2. For OCs with the highest precision, the range of [Fe/H] dispersion within OCs is reported to be between 0.04 dex and 0.08 dex \citep{jofre19}. However, it is important to note that the metallicity dispersion tends to increase with age. Therefore, caution should be exercised when using individual metallicities that exhibit a significant deviation from the mean value of OCs. 
To ensure data accuracy, we applied a data-cleaning process for each OC, excluding members with [Fe/H] values exceeding three standard deviations of the OC's mean metallicity. The final metallicity for each OC was then determined using the averaged values of remaining stars post-cleaning

\begin{figure}
   \centering
   \includegraphics[width=10cm]{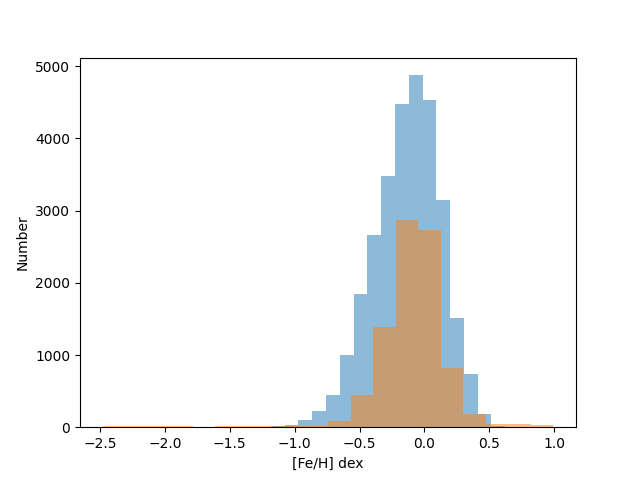}
      \caption{Histogram of the training set {and the metallicity that needed to be corrected for the LASP result}. The orange section represents the training data that were used to build the ANN model, while the blue section represents the metallicity that needed to be corrected for the selected samples.
 }
   \label{fighis}
   \end{figure}

In order to validate the applicability of the ANN model described in Sect. 2.2 to the current sample, it is important to have a training sample that covers a similar range as the parameters that need to be corrected. The input table contains the stellar metallicity values ranging from  -1 to 0.5 dex. Figure \ref{fighis} displays the overlapping portion of the histogram between the model's training data and the data that require correction. The plot demonstrates that the training set and the metallicity values to be corrected are comparable.

\begin{figure*}
   \centering
   \includegraphics[height=12cm]{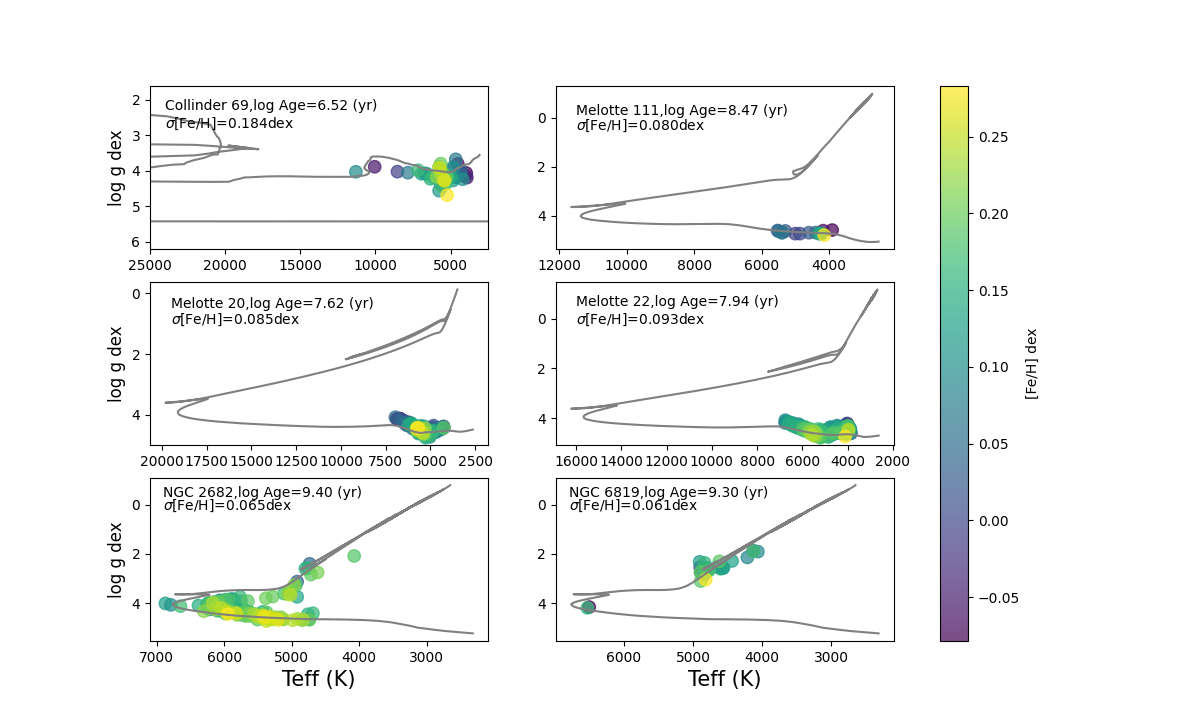}
      \caption{
 Kiel diagram displaying the member distribution for six selected OCs, color-coded according to metallicity. The x-axis corresponds to the effective temperature, while the y-axis corresponds to the surface gravity. The gray dashed track represents the PARSEC isochrone, which is based on the OC parameters from \cite{cavallo23}.
}
   \label{figcmd}
   \end{figure*}
\begin{figure*}
   \centering
   \includegraphics[height=10cm]{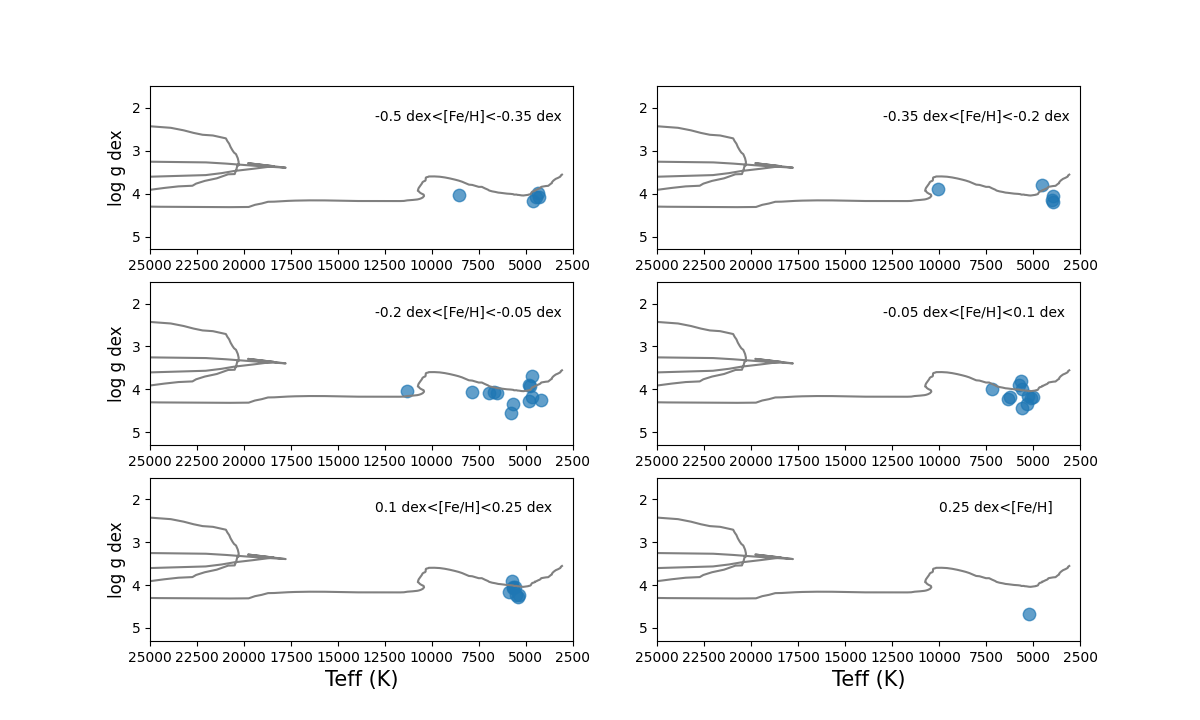}
      \caption{Kiel diagram and isochrone for Collinder 69, divided into six age bins.}
   \label{figcmd1}
   \end{figure*}

We verified the coverage of the input samples in the training set before eventually using the correction prediction. The final metallicity for each OC is checked by removing membership outliers, and even though they are in the multi-evolution phase, their final metallicity shows less dispersion. Figure \ref{figcmd} shows the member distribution of OCs on a color--magnitude diagrams (CMD) for six typical OCs in this study, with the color code indicating metallicity. The isochrone used is from the PARSEC V2.0 \footnote{http://stev.oapd.inaf.it/cgi-bin/cmd} isochrone \citep{nguyen22}, and the OC parameters are based on \cite{cavallo23} as shown in Fig. \ref{figcmd}. The members of the selected OCs span an age range from 130 Myr to 5.3 Gyr, and they are located in different evolution phases with a metallicity scatter of less than 0.2 dex. However, uncertainties and systematic errors in age measurements can cause a slight offset between targets and isochrones.
We examined whether or not there are variations in the fit differences when different metallicity bins are used, as shown in Fig. \ref{figcmd1} for Collinder 69 as an example. It was concluded that any potential systematic bias between the observational data and isochrones can be ignored.

\section{Comparison}\label{sec3}
\subsection{Metallicity comparison with  the literature for individual stars}
This section presents the iron abundances of the OCs as determined using high-resolution spectra mainly from three datasets: GALAH, APOGEE DR17 \citep{abdurro22}, and PASTEL \cite{soubiran16,soubiran22}. The last of these three is a collection of atmospheric parameters from high-resolution spectra (R > 25000) with S/N > 50. 
The PASTEL catalog comprises high-resolution data collected since the 1980s \citep{cayrel80, cayrel81, cayrel85, cayrel92, cayrel97, cayrel01} and is continually updated with recent high-resolution observations.

\begin{figure*}
   \centering
   \includegraphics[height=7cm]{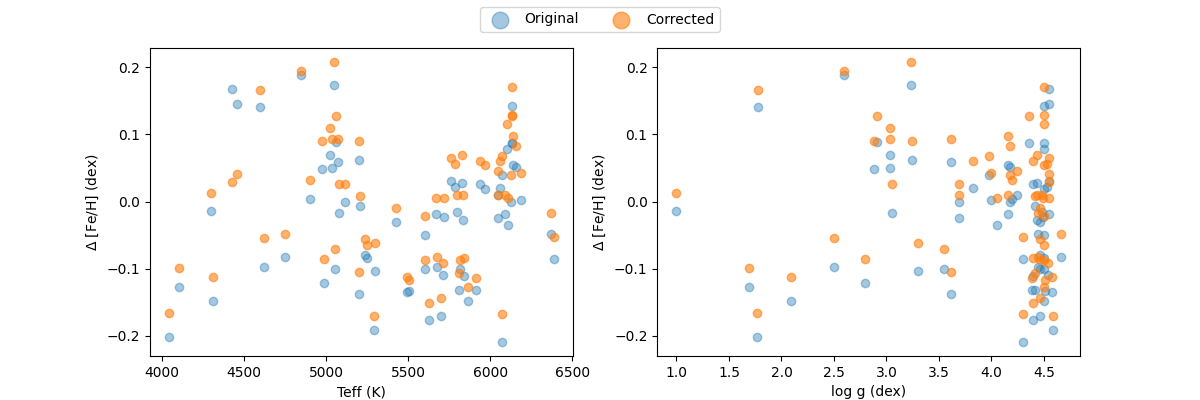}
      \caption{ [Fe/H] residual for 68 common stars between this work and that from the PASTEL catalog. }
   \label{figstd1}
   \end{figure*}

In this section, there are 68 stars in common with the PASTEL table, and the comparison results are shown in Fig. \ref{figstd1}. The average difference in metallicity is 0.000 dex, with a scatter of 0.093 dex for corrected [Fe/H] and LASP metallicity. The standard deviation for corrected [Fe/H] and the LASP metallicity is 0.096 dex. The difference in metallicity between PASTEL and other high-resolution surveys depends on the stellar iron ratio. However, for open clusters, the average deviation for both PASTEL and other surveys is approximately 0.04 dex. According to \cite{soubiran22}, the discrepancy and scatter are around -0.01$\pm$0.05 dex  for APOGEE,  0.00$\pm$0.07 dex for GALAH, and 0.02$\pm$0.05 dex for {\textit{Gaia}}-ESO. The dispersion we find is larger than what is reported in the literature. However, due to the limited number of samples that overlap, there are no noticeable patterns or biases between the corrected data and the original results.

\begin{figure*}
   \centering
   \includegraphics[height=7cm,width=21cm]{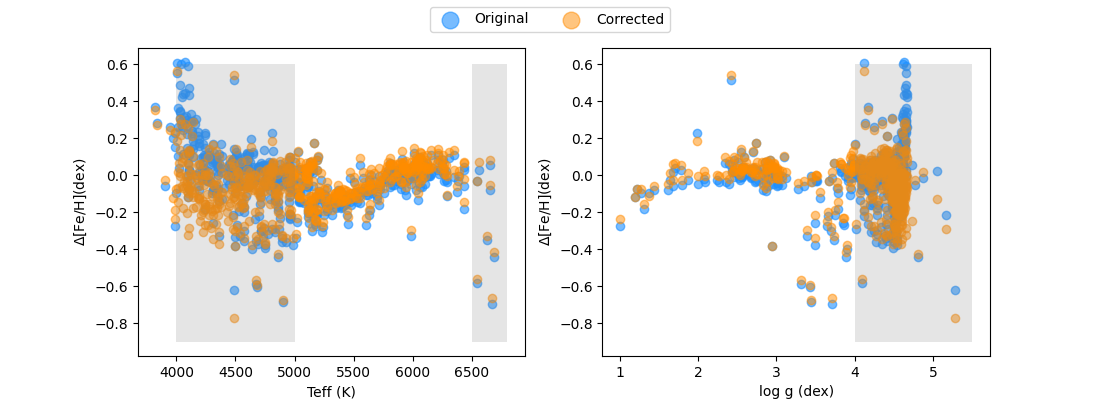}
      \caption{
Residuals between APOGEE DR17 and LAMOST LRS calculated based on the effective temperature and surface gravity. The blue dots represent the difference between HRS and the original LRS result, while the orange curve shows the difference between HRS and LRS after correction. The gray-shaded region includes samples with an effective temperature of below 5000 K or above 6500 K and dwarfs with a surface gravity of around 4.5 dex.}
   \label{figstd2}
   \end{figure*}

Additionally, a total of 592 cross-match samples were chosen between this study and APOGEE DR17 \citep{abdurro22}, and the comparison results are listed in Fig. \ref{figstd2}.
The average difference for this comparison is approximately -0.053 dex with a standard deviation of 0.140 dex for the original result, and -0.020 dex with an associated error of 0.160 dex for the corrected result. There is no systematic bias affecting our comparison of the metallicities provided by APOGEE DR17 and LAMOST LRS. However, there is an offset for stars with effective temperatures outside of a certain range (i.e., < 5000 K and 6500 K >) and for dwarfs with log g > 4.5 dex in the original results. This bias has been improved to some degree after correction.

\begin{figure*}
   \centering
   \includegraphics[height=7cm,width=21cm]{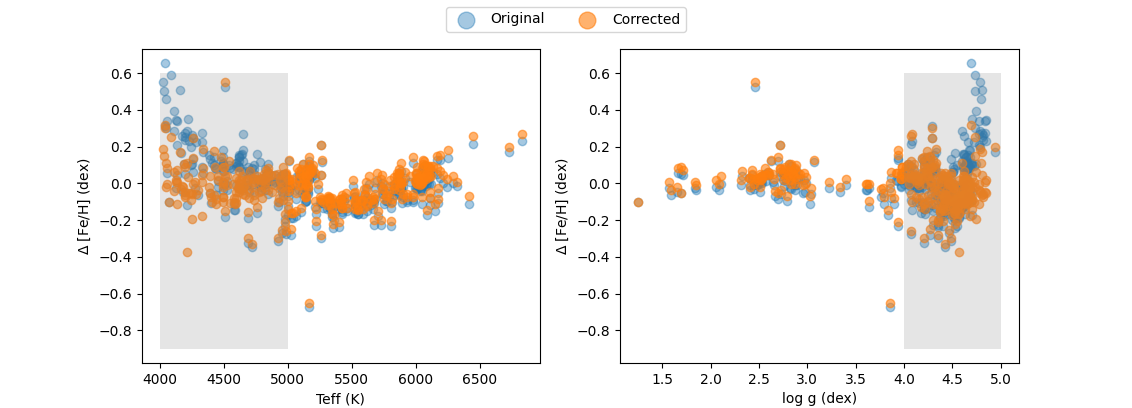}
      \caption{Residuals for metallicities between HRS \citep{spina21} and LRS with respect to effective temperature and surface gravity. The blue dots are the difference between HRS and the original result from LRS, and the orange circles show the difference between HRS and LRS after correction. The gray shadow highlights stars with T$_\textrm{eff}$ < 5000 K and log g > 4.5 dex. }
   \label{figstd}
   \end{figure*}

Furthermore, 453 stars were selected that are in common with the sample of \cite{spina21}. In Fig. \ref{figstd}, the [Fe/H] residual to effective temperature is shown. The blue point represents the difference between the literature value and the original value from the LAMOST catalog, while the orange point represents the difference between Spina's result and the [Fe/H] derived from LAMOST after correction. The blue point indicates a systematically lower offset compared to the GALAH result, especially for targets with temperatures cooler than 5000 K and with surface gravity of around 4.5 dex. The differences have been calibrated after the correction procedure described in Sect. 2.2, as shown by the orange dots in Fig. \ref{figstd}. Both systematic offsets are close to 0 dex before and after correction. However, for stars with $T_{\textrm{eff}}$ < 5000 K and log g ranging from 4.0 dex to 5.0 dex, the bias decreases in Fig. \ref{figstd}. Furthermore, the comparison indicates no systematic lower bias for hotter stars (above 6500 K).

\begin{figure*}
   \centering
   \includegraphics[height=12cm]{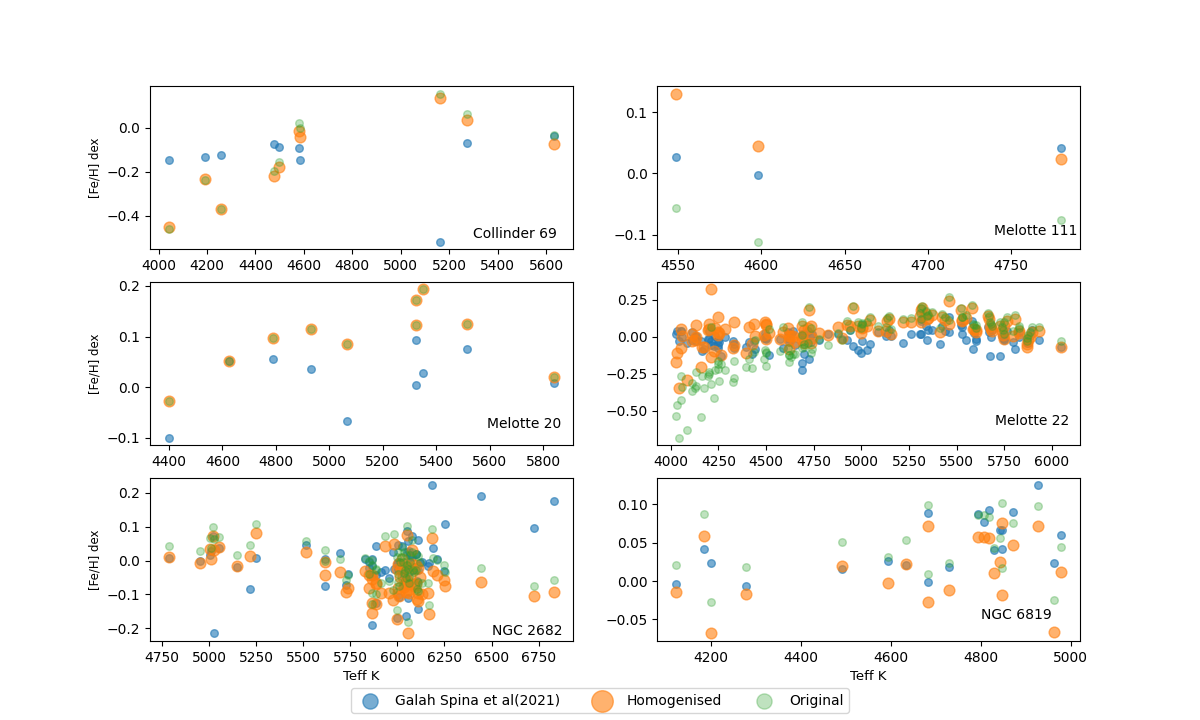}
      \caption{Membership metallicity distribution vs $T_{\textrm{eff}}$ and log g for six example OCs analyzed in the present study. The blue dots are the stars studied by \cite{spina21}. The green points are from LASP DR8 and the orange points are the metallicity for the present study after correction. }
   \label{figcluster}
   \end{figure*}

Additionally, we examined the iron abundance of the final members, and stars with [Fe/H] values that fall outside 3$\sigma$ of the standard deviation for OCs are excluded. Figure \ref{figcluster} displays the distribution of the [Fe/H] membership for six OCs (Collinder 69, Melotte 111, Melotte 20, Melotte 22, NGC 2682, and NGC 6819) that were studied in \cite{spina21} and selected as samples for internal comparisons within the OC. The blue, orange, and green points represent the iron abundance values from \cite{spina21}, the corrected values, and the original values, respectively. For all selected OCs samples, the [Fe/H] dispersion is decreased after correction.

\subsection{Comparison of OCs in terms of metallicity }
We looked for differences between the metallicities derived in the present study and those derived by other authors from the HRS of the OCs.
We include OC samples observed by APOGEE \citep{donor20}, {\textit{Gaia}}-ESO \citep{casali20}, and GALAH DR3\citep{spina21}. Additionally, we include data from the SPA project \citep{frasca19,casali20,dorazi20,paperI}, OCCASO \citep{casamiquela16}, OSTTA \citep{casamiquela16}, and the high-quality HRS results from \cite{netopil22}.

\begin{figure*}
   \centering
   \includegraphics[height=7cm]{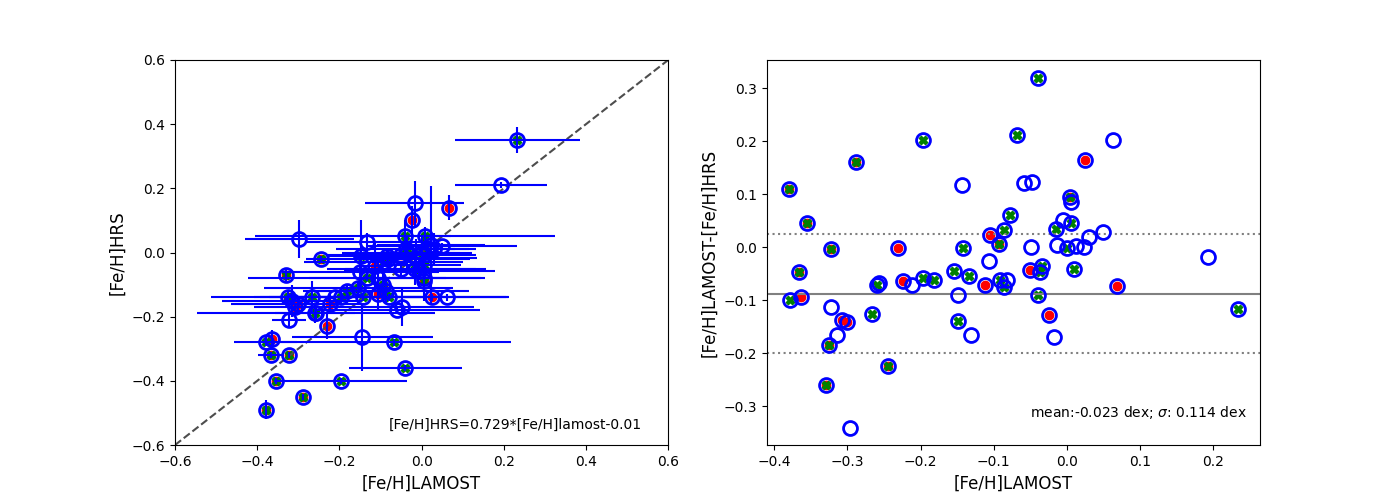}
      \caption{\textbf{Comparison of [Fe/H] Distribution between LAMOST and HRS Data. }Left panel:Distribution of [Fe/H] for both our results and results from the literature based on HRS. The metallicities from these two datasets follow the function $[Fe/H]_{HRS}=0.729[Fe/H]_{LAMOST}-0.01$. The dashed line represents a 1:1 linear relationship.
Right panel: Residual metallicities between the two datasets, with an average difference of -0.0023$\pm$0.114 dex. The red circles represent clusters with fewer than three members in the LAMOST data, while the green crosses represent HRS OCs with fewer than three members.
}
   \label{figcomp}
   \end{figure*}

A total of 88 samples have been selected for this comparison. Figure \ref{figcomp} shows the linear relationships between the metallicities derived in the present
study and those from other studies based on HRS. The bottom panel of Fig. \ref{figcomp} displays the residual metallicity of LAMOST's iron abundance. According to the literature, the average difference in this comparison is approximately -0.023 dex, with a dispersion of 0.114 dex. The red circles represent OCs with LAMOST membership of fewer than three stars, while symbols with green crosses represent HRS samples with fewer than three stars within a single OC.  
This suggests that the accuracy of average metallicity changes with the number of stars for each OC.


\subsection{Metallicity of OCs as derived with Monte Carlo sampling}

\begin{figure}
   \centering
   \includegraphics[height=8cm]{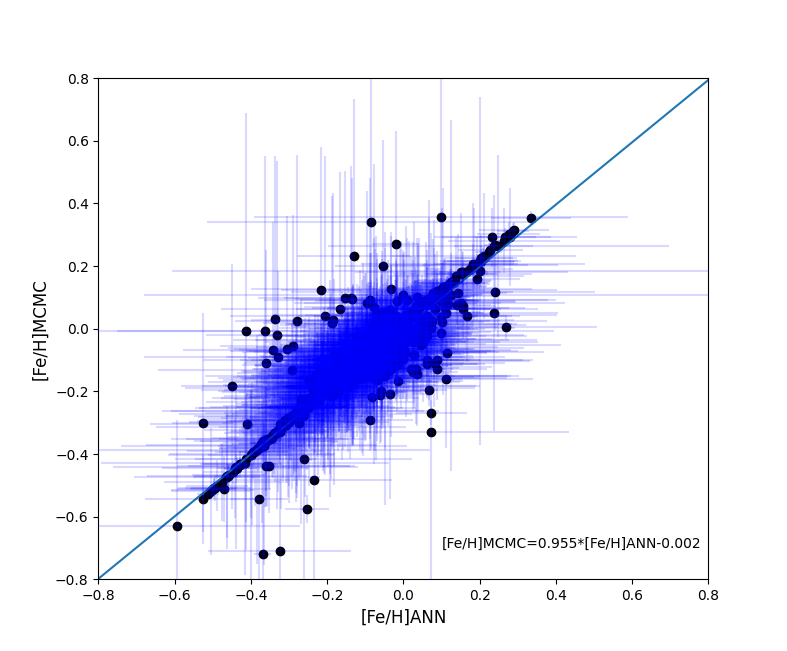}
      \caption{Linear regression relationship between OC metallicities determined with an ANN model  (x-axis)\ and an MCMC method (y-axis).}
   \label{fig_comp3}
   \end{figure}

In this section, we use the Monte Carlo  (MC) sampling method to perform parameter quality control and determine the metallicity of OCs for data comparison. The objective is to prevent any outliers in [Fe/H] within a single OC. This is because there is a slight dispersion (within 0.1 dex) in the metallicity of OC memberships, even in different evolution phases \citep{bertelli18,semenova20}. The method primarily focuses on the accuracy of the stellar parameters, membership selection, evolution stage checking, and data cuts, which are discussed in the previous section. The basic parameter assumptions are based on \cite{fu22}. The uncertainties in [Fe/H] follow a Gaussian distribution, with a mean uncertainty of 0.07 dex, which is the typical error for the LASP pipeline. For each cluster, we randomly selected 5000 samples to determine the mean [Fe/H]$_{mean}$ and median [Fe/H]$_{med}$ values, finding errors of around 0.12 dex and 0.15 dex, respectively. The mean value is used as the OC metallicity, and the standard deviation is the final uncertainty. Figure \ref{fig_comp3} shows the correlation between OC metallicity based on ANN and MC. Generally, there is good agreement between the OC metallicity obtained from the two methods, with a systematic bias of approximately -0.005 dex and a standard deviation of 0.064 dex. However, the [Fe/H] derived from ANN is used for further discussion for each OC.


%

%

\section{Discussion}\label{sec4}

Most of the targets observed by LAMOST are located in the anti-Galactic center direction, around a galactocentric distance of 8 kpc, ranging from areas near the Sun to farther regions, but with a lack of samples between the Galactic center and the Sun.These samples are ideal for studying the chemical properties and migration patterns of the solar neighborhood. In a study by \cite{schonrich09}, the authors discussed the migration of stars from the inner disk to the vicinity of the Sun in the Galactic disk.
Galactic disk. Their model suggests that this migration is linked to the formation of the thin and thick disks. This section of the discussion focuses on the chemical properties as they vary with galactocentric distance (R$_{GC}$) and guiding distance (R$_{guid}$). We will discuss a method using a semi-empirical function to estimate the birthplaces of OCs later, particularly for samples located in the solar neighborhood.

\subsection{Guiding radii of LAMOST OCs}

We used {\textit{Gaia}} DR3 and the reference catalog to gather photometric information about OCs, which includes some basic details. However, instead of using the current galactocentric radius, we can obtain more accurate metallicity gradients by using the guiding center radius of a cluster, which is not affected by orbital blurring \citep{netopil22,spina21,zhang21}.
The guiding distance is calculated as R$_{guiding}$= L$_{z}$/V$_{c}$(R) where L$_{z}$ represents the angular momentum in the z-axis in the Galactic coordinate system, and V$_{c}$(R) represents the circular velocity at the cylindrical radius R. The Cavallo catalog does not provide the guiding distance of the OCs. However, in a study by \cite{myers22}, the authors calculated the guiding distance using the Milky Way model mentioned by \cite{price21}. Additionally, \cite{spina21} discuss the distribution of [Fe/H]--R$_{guiding}$ and calculate R$_{guiding}$ using the GALPY package\footnote{https://github.com/jobovy/galpy}. In this section, we present the
dynamical properties of our sample of OCs as calculated using the GALPY package \citep{bovy15}. The inputs for the calculation of  
R$_{guiding}$ include observational parameters such as coordinates, distance, proper motions, and RVs.The work by \cite{spina21} discusses the angular momenta \(J_{R}\), \(L_{Z}\), and \(J_{Z}\), as well as the guiding radius \(R_{guid}\) with respect to the MWpotential2014 \cite{bovy15}. The information for the current samples of OCs, including their metallicities and spatial distributions, can be found in Table \ref{tab:A1}.


\subsection{Metallicity distribution across Galactic space}
This section examines the metallicity properties across various regions of the Galaxy's spatial distribution. We discuss the relationship between the metallicity distribution [Fe/H]--R$_{GC}$ with Galactocentric distance and the spatial distribution of metallicity abundance with cluster guiding radius [Fe/H]--R$_{guid}$.
The model and the observational data indicate that the distribution of metallicity shows a decreasing trend with increasing distance from the Galactic center. In this section, we discuss the characteristics of metallicity as it changes with galactocentric distance. The age and galactocentric distance of the OC samples in this section are based on the catalog provided by \cite{cavallo23},  which contains information for over 1000 OCs younger than 1 Gyr, with the oldest OC being 9 Gyr old. These OCs cover a wide range of galactocentric distances (R$_{GC}$) from 4 kpc to 16 kpc, with more than 100 OCs showing a R$_{GC}$ of greater than 11 kpc.

\begin{figure}
   \centering
   \includegraphics[height=7.5cm]{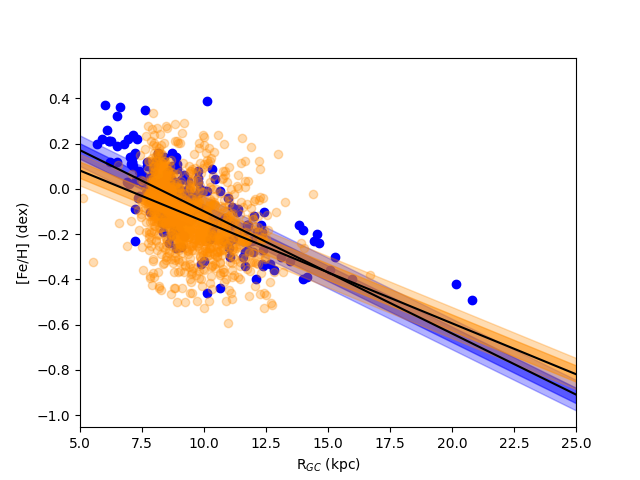}
      \caption{
Metallicity distribution as a function of galactocentric distance for OCs of all ages. The orange dots represent the sample from this work, while the blue points represent OCs selected from the literature where HRS was used to determine metallicity. The linear regression models for the present samples are shown in dark orange and orange, representing the 68\% and 95\% confidence intervals, respectively. Similarly, the confidence intervals for the literature samples are shown in dark blue and blue, representing the 68\% and 95\% confidence intervals, respectively.
}
   \label{fig_Dis}
   \end{figure}

For clarity, the error associated with the OC metallicity is not plotted in Figure \ref{fig_Dis} and the subsequent figures, which shows the metallicity distribution versus R$_{GC}$ for all age and distance ranges compared with a similar distribution based on the literature. Despite the presence of over 1,000 OCs in this study, only a limited number of these clusters are located within 8 kpc. This scarcity contributes to discrepancies between our analysis results and those reported in the literature. In this section, the Bayesian linear regression between metallicity and galactocentric distance with {\sc pymc3} package \cite{2016ascl.soft10016S}, adopted a similar estimate to \cite{spina21} and \cite{cavallo23}. A simple linear regression relationship is used in the present fitting: $Y_{i}=\alpha * X_{i} + \beta$. Here,$Y_{i}$ and $X_{i}$ are R$_{GC}$ and the OC metallicity following a normal distribution:
\begin{equation} 
X_{i}: N(R_{GCi},\sigma_{R_{GC},i}), \\
Y_{i}: N([Fe/H]_{i}, \sigma_{[Fe/H],i}),
 \label{equ3}
\end{equation}
and the values are R$_{GC}$ and [Fe/H] for each OC, which have the maximum possibility in the normal distribution. The $\sigma_{R_{GC}, i}$ is the calculated standard deviation based on the R$_{GC}$ estimated using the ANN model \citep{cavallo23}, and the dispersion of [Fe/H] is derived in the metallicity calculation for each OC. 
 The priors for $\alpha$ and $\beta$ follow the distributions $N(-0.068 \textrm{ dex/kpc}, 0.1 \textrm{ dex/kpc})$ and $N(0.5 \textrm{ dex}, 1 \textrm{ dex})$, respectively. The prior free parameter $\epsilon$ follows the positive part of the half-Cauchy distribution, where $\gamma$ is 1, based on the assumptions outlined in \cite{spina21}. This study utilizes 10,000 samples in the simulation, employing the methods mentioned above.The final parameters for the linear regression are listed in Table \ref{tab-grad-noi}, and we employed the same fitting method as reported in the literature. We show the gradient of [Fe/H]--R$_{GC}$ from other work in Table \ref{tab-grad}; the trend is around -0.060 dex~kpc$^{-1}$ \citep{jacobson16,donor20,spina21,paperI,myers22}. The metallicity gradient is relatively flat in the present calculation, with a value of $-0.045\pm0.006$ dex~kpc$^{-1}$; however, the trend from these latter works is based on HRS, and the measurement we obtain in the present study is less precise. The trend increases to $-0.054\pm0.004$ dex~kpc$^{-1}$ in the studies that use the same fitting method as that employed here. 
Moreover, the metallicity trend derived from \cite{friel02}, \cite{reddy16}, \cite{carrera19}, \cite{casamiquela19}, and \cite{cavallo23} is comparable with this measurement considering the uncertainties. We adopted the same Bayesian fitting method to fit the linear regression parameters of [Fe/H] and OC guiding distance. The parameters of the linear relationship are listed in Table. \ref{tab-grad-noi}.

\begin{figure}
   \centering
   \includegraphics[height=7.5cm]{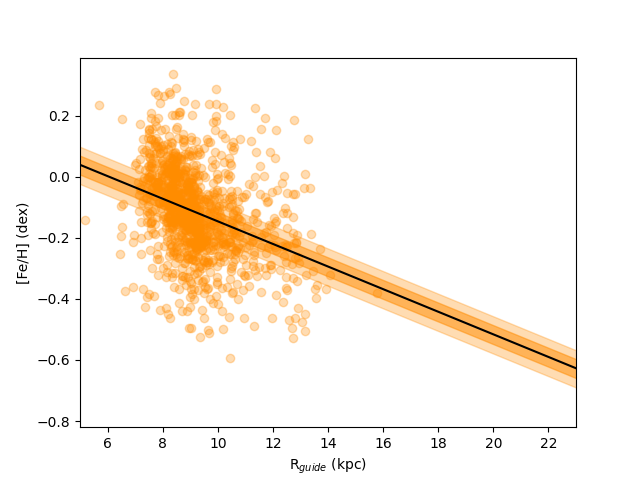}
      \caption{Metallicity distribution as a function of guiding distance for our sample of OCs. The dark and light orange areas are the  68\% and 95\% confidence intervals for our sample, respectively.}
   \label{fig_Dis2}
   \end{figure}

\begin{figure}
   \centering
   \includegraphics[width=9.3cm]{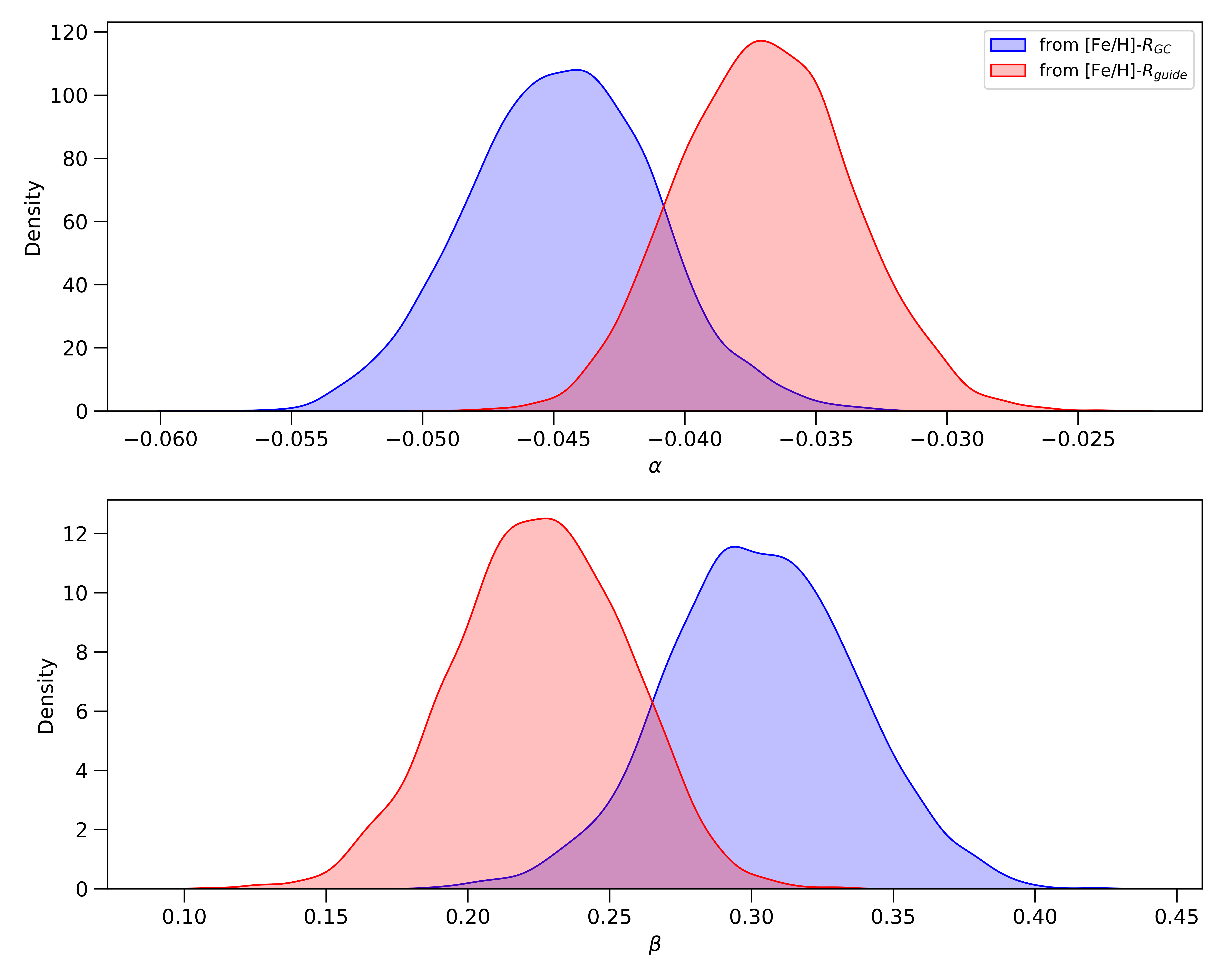}
      \caption{Posterior distributions of parameters $\alpha$
and $\beta$ obtained using Bayesian linear regression analyses of   [Fe/H]--R$_{GC}$ and [Fe/H]--R$_{guide}$.}
   \label{fig_normal}
\end{figure}

 Figures \ref{fig_Dis} and  \ref{fig_Dis2} show the metallicity distribution as a function of either distance from galactic center or guiding distance for the present samples, respectively. In Fig. \ref{fig_Dis}, the orange dots represent the OCs
from the present work, while the blue points represent OCs selected from the literature
where HRS were used to derive metallicity. The light shadow and dark shadow are the 95$\%$ and 68$\%$ confidence intervals for the final Bayesian regression. In this calculation, only 111 OCs contain more than three stars, which may lead to uncertainties in metallicity determination.  We used the MCMC method mentioned in \cite{spina21} to determine that the metallicity gradient flattens as the Galactocentric distance reaches 9.129 kpc. The metallicity gradient in the inner disk is $-0.041 \pm 0.078$ dex~kpc$^{-1}$. However, the processes of churning and blurring can lead to significant dispersion in the samples, as noted by \cite{spina21}.
In contrast, the outer disk shows a metallicity trend of $-0.029\pm0.013$ dex~kpc$^{-1}$. \cite{donor20} mention a  “Keen” at around 12 kpc, but the limited number of OCs within 7 kpc may account for the observed difference.
All of the samples mentioned here are situated within 1 kpc from the Galactic midplane. The metallicity trend, which can be observed in Fig. \ref{fig_Dis2}, follows the gradient [Fe/H]--R$_{guide}$, corresponding to R$_{GC}$. The value of this trend is $-0.037\pm0.003$ dex~kpc$^{-1}$, suggesting that the gradient is slightly flatter than the [Fe/H]--R$_{GC}$ relationship, but the two generally align. We observed a change in the trend of [Fe/H]--R$_{guide}$ at a guiding distance of approximately 9.518 kpc. At this distance, the gradient will shift from $-0.062\pm0.041$ to $-0.025\pm0.011$. 
To evaluate the alignment between our fitted results and the observed data, we conducted a Kolmogorov-Smirnov (KS) test, considering scenarios with and without metallicity uncertainties. We note that the maximum discrepancy between the cumulative distribution functions (CDFs) of the fitted and observed data is approximately 29\%. Additionally, the inclusion or exclusion of [Fe/H]-related uncertainties appears to have little impact on the results. We conducted a similar comparison in \cite{spina21} for the posterior distributions of fitting parameters shown in Fig. \ref{fig_normal}. However, as opposed to these latter authors, who reported an identical posterior distribution for $\alpha$ and $\beta$, we find a clear difference in the posterior distribution.

 \begin{table*}[ht]
\setlength{\tabcolsep}{1.25mm}
\begin{center}
\caption{Linear regression parameters and the 95\% confidence intervals. The left panel shows the fitting results for the present sample and the left panel shows the linear regression parameters for selected samples from previous works.} 
\begin{tabular}{lrrrr|rrrr}
\hline\hline
&Parameter & Mean & Uncertainty& 95\% C.I & Mean&Uncertainty&95\% C.I\\
\hline
[Fe/H]-R$_{GC}$ &$\alpha$&-0.045 & 0.006 &[-0.052,-0.037] &-0.054&0.004&[-0.060,-0.046] \\
&$\beta$& 0.304 & 0.033&[0.240,0.378] &0.441&0.034&[0.371,0.506]   \\
\hline
[Fe/H]-R$_{guiding}$ &$\alpha$&-0.037 & 0.003  &[-0.043,-0.030] &   & \\
&$\beta$& 0.224 & 0.031 &[0.160,0.282] &   & \\
 \hline
\end{tabular}
\label{tab-grad-noi}
\end{center}
\end{table*}

\begin{table*}[ht]
\setlength{\tabcolsep}{1.25mm}
\begin{center}
\caption{Slope of the metallicity gradient from selected literature papers.
}
\begin{tabular}{llccl}
\hline\hline
\multicolumn{1}{c}{Reference}&\multicolumn{1}{c}{sample}&\multicolumn{1}{c}{d[Fe/H]/d R$_{gc}$}&\multicolumn{1}{c}{N OCs} &\multicolumn{1}{c}{Comment}\\
\hline
\cite{friel02}  &low-res optical spectra &-0.059$\pm 0.010$&39 &7<R$_{gc}$<16 kpc, all ages$^a$ \\
\cite{reddy16} &high-res optical spectra &-0.052$\pm 0.011$&67 &6<R$_{gc}$<12 kpc and $|z|<$500 pc\\
               &&-0.015$\pm 0.007$&12 &12<R$_{gc}$>24 kpc$^b$ \\
\cite{carrera19} &APOGEE DR14, GALAH DR2&-0.052$\pm 0.003$&46 &6<R$_{gc}$<13 kpc$^c$ \\
                 &&-0.077$\pm 0.007$&  &6<R$_{gc}$<11kpc$^c$ \\
                 &&0.018$\pm 0.009$&  &11<R$_{gc}$<13 kpc$^c$ \\
\cite{casamiquela19} &high-res optical spectra &-0.056$\pm 0.011$&18 &R$_{gc}$<12 kpc, all ages$^d$ \\
\cite{donor20} &APOGEE DR16&-0.068$\pm 0.004$&68&R$_{gc}$< 13.9 kpc, all ages$^e$  \\
               &&-0.009$\pm 0.011$&3&R$_{gc}$>13.9  kpc, all ages\\
\cite{myers22} &APOGEE DR17&-0.055$\pm 0.001$& & 6<R$_{gc}$<11.5 kpc, all ages\\              
\cite{spina21} &high-res optical spectras&-0.076$\pm 0.009$&134& 5<R$_{gc}$<12 kpc, all ages\\
\cite{paperI} &high-res optical spectras&-0.066$\pm 0.005$& & R$_{gc}$<14 kpc, all ages\\
\cite{cavallo23} & photometric data &-0.048$\pm 0.009$& & 5<R$_{gc}$<12 kpc,ages > 500 Myr  \\
\hline
\end{tabular}
\label{tab-grad}
\end{center}
\end{table*}
\subsection{Metallicity distribution compared with a chemo-dynamical model}

The chemical properties we presently observe are the outcome of intricate evolutionary processes, disk asymmetries, and non-equilibrium processes, as mentioned in studies of the {\textit{Gaia}} dataset, such as \cite{wang2023a} and \cite{2018ApJ...865...96F}. For instance, according to \cite{sellwood02}, migration within the disk is a crucial process that needs to be taken into account when modeling galactic evolution. In their work, \cite{minchev13, Minchev14a, Minchev14b} considered the impact of migration and kinematic heating on galactic disk modeling, but did not address the issues related to self-consistent simulation.


In this section, we discuss the predictions based on chemo-dynamical models   \citep[][e.g. the MCM model]{Minchev14a} and compare these to observational findings. \cite{Minchev14a} provide the galactic metallicity distribution based on the galactocentric distance split into 17 age bins. MCM takes into account the distance from the disk midplane and migration, combining chemical evolution and dynamics.

{Figures \ref{fig_model1} and \ref{fig_model2}} show observational results compared with predictions split by age and color-coded according to the number of OCs in each cluster. The chemo-dynamic track is separated into $|z|<0.3$ kpc and 0.3 kpc $<|z|<$ 0.8 kpc \citep{Minchev14a}. In the comparison plots, the red line represents the predicted track with $|z|<0.3$ kpc, while the blue line represents the predicted track for 0.3 kpc $<|z|<$ 0.8 kpc. According to the model, the midplane distance of OCs for these samples is limited to 0.8 kpc. The cross symbols represent OCs with fewer than three members. As the plot shows, most of the OCs of the compared sample are younger than 0.6 Gyr. The metallicity distribution is systematically lower for the youngest age bins (0.0 < age < 0.3 Gyr and 0.3 < age < 0.6 Gyr), and OCs with fewer than three members show a greater dispersion compared to the chemo-dynamic prediction. The observational data are consistent with the predictions for ages of greater than 1.2 Gyr, as mentioned in previous work \citep{paperI}.


\subsection{Birth radii of open clusters}
\cite{sellwood02} first suggested that transient spiral arms can permanently change the angular momentum of stars without heating their orbits, causing them to migrate away from their birth location over time.
Moreover, due to the exponential density profile of the Galaxy, a population of field stars will preferably move from the inner disk to the outer disk. 
\cite{reddy16} suggest that OCs follow a similar trend to field stars, in that those stars currently in the outer disk were born within 12 kpc and have migrated. 
This process can flatten the birth metallicity gradient over time \citep{minchev13}; as a result, radial
migration has to be taken into account when trying to understand the birth environment of OCs  \citep{minchev18, ratcliffe22}.
The present work adopted an empirical method from \cite{lu22a} to infer the birth radii of their OC sample.
This method is built upon the previous method developed by \cite{minchev18} and the assumption that the ISM metallicity gradient (which is a function of lookback time) has a linear relationship with the metallicity range in mono-age populations. 

\begin{figure}
   \centering
   \includegraphics[width=10cm]{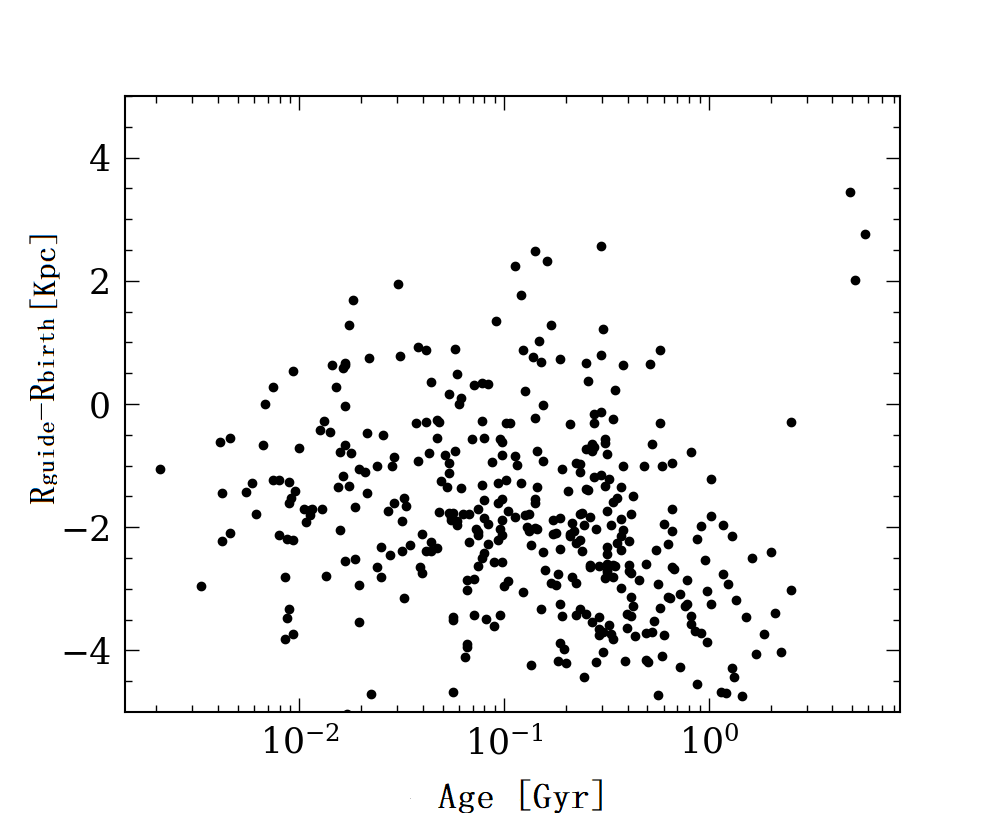}
      \caption{ Migration distance for the present sample of OCs as a function of their age. }
   \label{figRg}
   \end{figure}

For the present study, we used age and metallicity measurements to estimate the birth radii for the OCs of our sample following the method described in \cite{lu22a}. 
The birth radii, $\mathrm{R_b(age, [Fe/H])}$, are calculated using 
\begin{equation}
    \mathrm{R_b(age, [Fe/H])} = \frac{\mathrm{[Fe/H]} - \mathrm{[Fe/H]}(0,\tau)}{\mathrm{\Delta [Fe/H](\tau)}}
,\end{equation}
where [Fe/H] is the observed metallicity, and [Fe/H] (0,$\tau$) and $\Delta$[Fe/H] ($\tau$) represent the metallicity at the Galactic center and the metallicity gradient at a given lookback time, $\tau$, respectively. 
$\Delta$[Fe/H] ($\tau$) and [Fe/H] (0,$\tau$) are taken from Table A1 in \cite{lu22a}. 
Figure \ref{figRg} shows the migration distance for the OCs in this study as a function of age. 
Surprisingly, according to this analysis, most OCs have migrated from the outer disk to their current guiding radius. 
However, the oldest OCs have migrated from the inner Galaxy.
This trend, if real, could suggest that the youngest clusters most likely cannot survive the tidal disruption of the inner Galaxy. 
However, as the relation used to infer birth radii is only calibrated with field stars of $>$ 1.5 Gyr old \citep{xiang22}, inferring birth radii for these young clusters can be biased.
As a result, we do not draw any conclusions on this topic here, and propose that further testing  is necessary, with particular attention to any possible bias that might affect the derived birth radii of young field stars.

\section{Summary}\label{sec5}

This study uses low-resolution spectra obtained from the LAMOST to re-evaluate the metallicity abundance of over 1131 open clusters (OCs) identified in the {\textit{Gaia}} DR3 catalog \citep{cavallo23}. The selected OCs in this study span an age range from a few million years to 9.5 Gyr, with the majority of OCs being younger than 600 Myr. These samples are primarily located in the solar neighborhood, covering a galactocentric distance ranging from 4.68 kpc to 16.7 kpc. The key findings of this study are summarized as follows:

\begin{enumerate}
      \item 
       We used the LASP pipeline in the LAMOST LRS DR8 to determine the average RV for a set of selected OCs. The mean difference between the RVs obtained from the LASP measurement and those derived from the HRS measurement is $5.598 \pm 14.031$ km/s. Additionally, there is an offset of $6.075 \pm 11.496$ km/s between our results and the determination from {\textit{Gaia}} DR3.

      \item 
The stars with a temperature of below 5000 K and the dwarfs with a surface gravity of around 4.5 dex consistently exhibit lower [Fe/H] values compared to the original LAMOST DR8 catalog. To determine the metallicity of the OCs, we used an ANN that incorporates the correction method introduced by \cite{niu23}. 
This approach helped us re-identify the metallicity of the stars and minimize bias in the process. Consequently, the discrepancy in metallicity between our work and the HRS result is compatible after corrections are made.

       
      \item 
We examined the distribution of metallicity across the spatial extent of the galaxy. The metallicity gradient, determined through Bayesian linear regression, is calculated as $-0.0045*R_{GC} + 0.304$ km~dex$^{-1}$. This gradient is relatively flat. Notably, this trend is consistent with the results obtained from literature samples and the previous calculation presented in Table \ref{tab-grad}.

      \item 
We compared the observed trend of metallicity with the chemo-dynamical predictions \citep{minchev13, Minchev14a, Minchev14b} based on age. The results from LAMOST indicate a lower metallicity compared to the predictions when the observational uncertainties are taken into account. Moreover, the metallicities of  observed OCs  are not taken into account.
Furthermore, the metallicities of observed OCs in the youngest age groups show a larger dispersion, which can be attributed to the fact that most OCs currently have a limited number of members ---typically around three stars. This large scatter is also present in the youngest age groups of the HRS \citep{paperI}, and is likely due to the measurement uncertainties associated with young stars.

      \item 
We determined the birth radii of our sample of OCs using the method presented by \cite{lu22a}. Our findings suggest that the majority of the OCs near the Sun originate from the outer disk, indicating that the youngest OCs may have been destroyed in the inner disk as a result of tidal interaction. However, additional testing is required to confirm these findings.

   \end{enumerate}

\begin{acknowledgements}
     We would like to thank the anonymous referee for his/her very helpful and insightful comments and Ivan Minchev for sharing his evolutionary models. GJW acknowledges the Africa Europe Cluster of Research Excellence (CoRE-AI) fellowship. HFW is supported in this work by the Department of Physics and Astronomy of Padova University though the 2022 ARPE grant: Rediscovering our Galaxy with machines. JXC acknowledges the Support of the Postdoctoral Fellowship Program of CPSF under Grant Number GZC20240124. This work exploits the Simbad, Vizier, and NASA-ADS databases and the software TOPCAT \citep{topcat}.  
     
     The Guo Shou Jing Telescope (the Large Sky Area Multi-Object Fiber Spectroscopic Telescope, LAMOST) is a National Major Scientific Project built by the Chinese Academy of Sciences. Funding for the project has been provided by the National Development and Reform Commission. LAMOST is operated and managed by National Astronomical Observatories, Chinese Academy of Sciences. 
     This work has made use of data from the European Space Agency (ESA) mission {\textit{Gaia}} (https://www.cosmos.esa.int/Gaia), processed by
the Gaia Data Processing and Analysis Consortium (DPAC,
https://www.cosmos.esa.int/web/Gaia/dpac/consortium). This work is supported by the National Natural Science Foundation of China under program Nos. 12090040, 12090043, and 12003025, as well as the Basic Research Program of Yunnan Province (No. 202401AT070142). We also express our gratitude for the generous support from the International Centre of Supernovae, Yunnan Key Laboratory (No. 202302AN360001), and the Natural Science Foundation of Yunnan Province (No. 202201BC070003).
\end{acknowledgements}

%
%

\begin{appendix}
\section{The physics information for Individual OCs}    
\renewcommand{\thetable}{A\arabic{table}}
\onecolumn
\begin{table*}[ht]
\setlength{\tabcolsep}{1.25mm}
\begin{center}
\caption{Basic information of OCs in the present samples (extract). The full table is available at the CDS.}
\begin{tabular}{lccccccccccc}
\label{tab:A1}\\
\hline\hline
Name        &  RA &  Dec&  R$_{GC}$ &   L$_{z}$ &  R$_{guiding}$& R$_{birth}$& [Fe/H]$_{ANN}$ & $\sigma_{[Fe/H],ANN}$  & [Fe/H]$_{MC}$ &  $\sigma_{[Fe/H],MC}$ &  N     \\
             & (J2000)   &(J2000)   & (kpc) & (kpc)  & (kpc)  & (kpc)  & (dex)     &(dex)  &(dex)&(dex)  \\
\hline
ADS 16795  & 23:30:22.12 & 58:33:12.29 & 8.199 & 1719.15 & 7.79 & 13.31 & -0.253 & 0.057 & -0.574 & 0.044 & 1  \\
ASCC 100   & 19:01:36.74 & 33:36:03.40 & 7.963 & 1723.86 & 7.82 & 9.89 & -0.017 & 0.111 & -0.00014 & 0.163 & 1  \\
ASCC 101   & 19:13:22.73 & 36:21:35.64 & 7.981 & 1705.06 & 7.72 & 10.71 & -0.062 & 0.084 & -0.020 & 0.075 & 1  \\
ASCC 105   & 19:41:53.18 & 27:23:09.94 & 7.872 & 1684.99 & 7.62 & 11.29 & -0.111 & 0.165 & -0.087 & 0.165 & 1  \\
ASCC 108   & 19:53:32.93 & 39:20:33.36 & 7.887 & 1723.58 & 7.82 & 10.70 & -0.045 & 0.262 & -0.022 & 0.116 & 2  \\
ASCC 11    & 03:32:12.10 & 44:50:26.76 & 8.888 & 1878.42 & 8.60 & 13.91 & -0.266 & 0.142 & -0.254 & 0.157 & 2  \\
ASCC 113   & 21:11:51.02 & 38:32:43.09 & 8.073 & 1786.69 & 8.14 & 10.86 & -0.073 & 0.057 & -0.039 & 0.071 & 6  \\
ASCC 12    & 04:49:40.94 & 41:42:51.48 & 9.138 & 1949.23 & 8.97 & 12.36 & -0.151 & 0.142 & -0.126 & 0.152 & 2  \\
ASCC 123   & 22:42:15.05 & 54:11:56.09 & 8.268 & 1788.35 & 8.14 & 9.48 & 0.068 & 0.099 & 0.095 & 0.030 & 1  \\
ASCC 128   & 23:20:38.75 & 54:34:24.10 & 8.37 & 1795.31 & 8.18 & 11.75 & -0.131 & 0.260 & -0.116 & 0.254 & 1  \\
...........\\
\hline
\hline
\end{tabular}
\end{center}
\end{table*}
\section{The comparison between and chemo-dynamic model}  
\renewcommand{\thefigure}{B\arabic{figure}}
\begin{figure*}
   \centering
   \includegraphics[width=17cm]{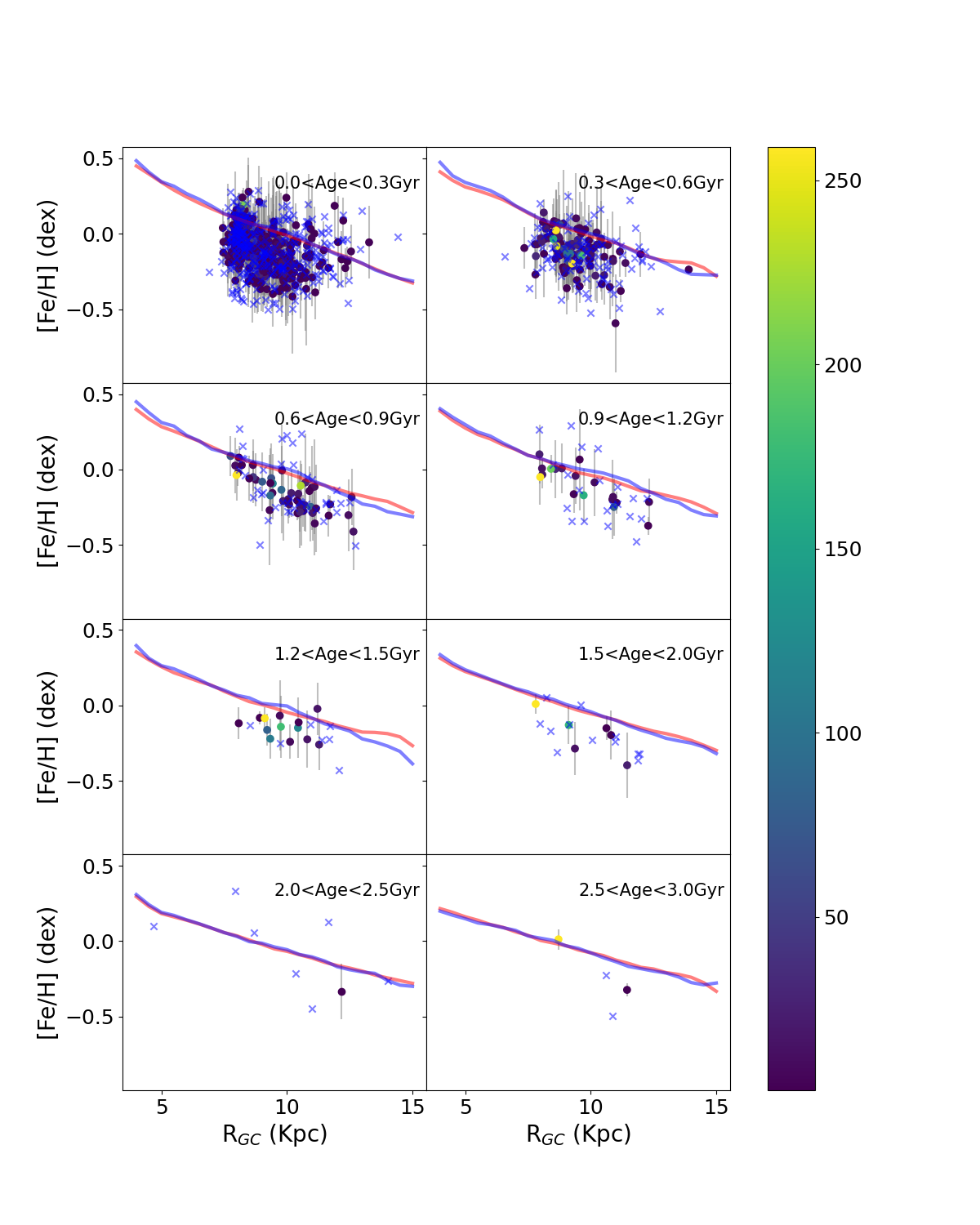}
      \caption{Comparison between chemo-dynamic model \citep{Minchev14a,Minchev14b} and observation for clusters younger than 3Gyr. The red lines and the blue lines are predictions from the MCM models for $|z| < 0.3$\,kpc and 0.3 $< |z| < $0.8 kpc respectively. The colors in the cycle are the number of stars for each OC whose membership is larger than 3, and the cross symbol is the OC with a member less than 3.} 
         \label{fig_model1}
   \end{figure*}

   \begin{figure*}
   \centering
   \includegraphics[height=20cm,width=17cm]{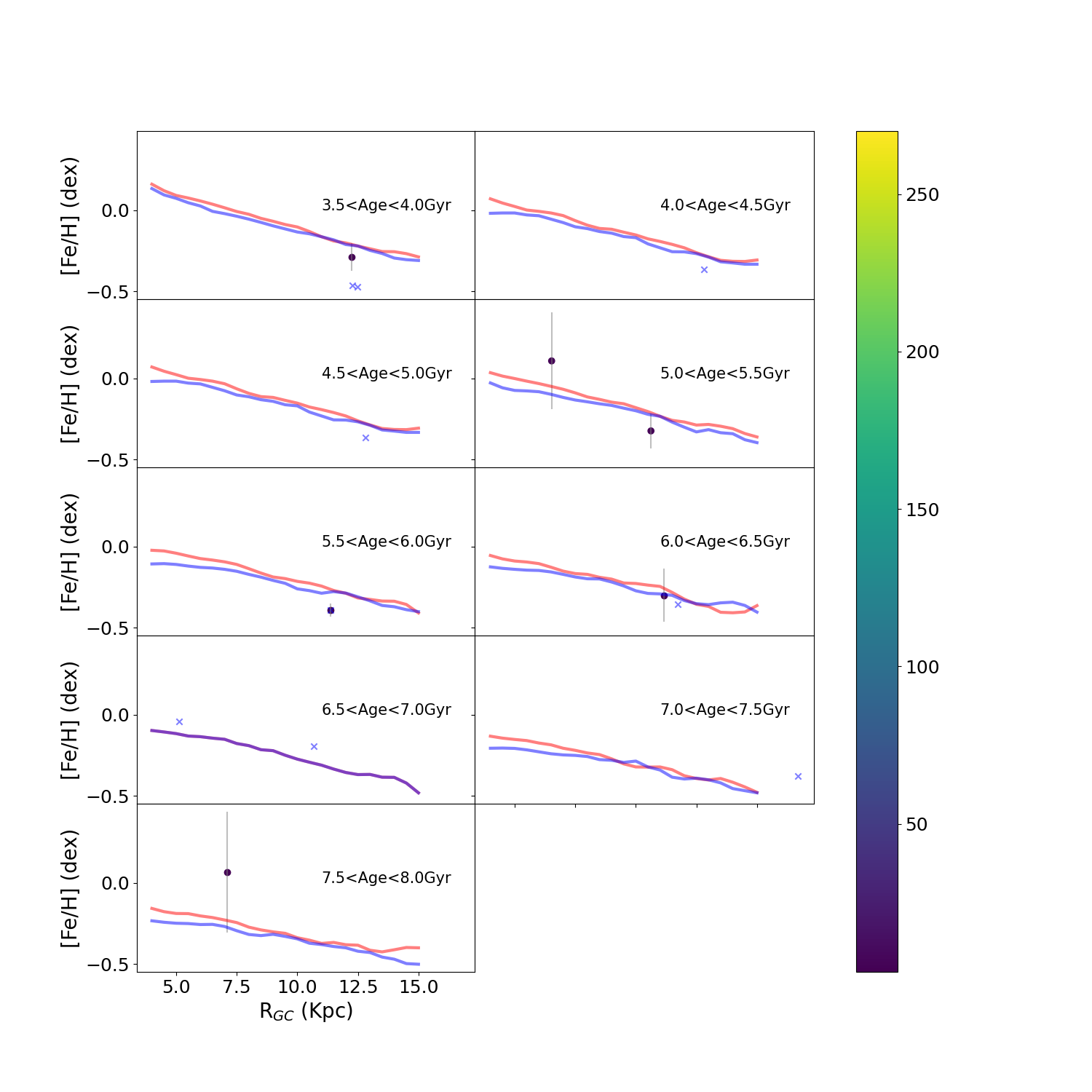}
      \caption{Same as Fig.~\ref{fig_model1} but for clusters between 1.5 and 4.5 Gyr. Note the paucity of OCs older than 2.5 Gyr} 
         \label{fig_model2}
   \end{figure*}
\end{appendix}

\end{document}